\input{aipcheck}

\documentclass[
    ,final            
  ]
  {aipproc}

\layoutstyle{6x9}

\usepackage{mathrsfs}
\usepackage{amssymb}
\usepackage{color}
\usepackage{graphicx}

\newcommand{\B}{{\mathcal B}}

\begin{document}

\title{On the definition of equilibrium and non-equilibrium states in 
dynamical systems}


\author{Takuma Akimoto}{
  address={Department of Applied Physics, Advanced School of Science and Engineering, Waseda University, Okubo 3-4-1, Shinjuku-ku, Tokyo 169-8555, Japan.}
}



\begin{abstract}
We propose a definition of equilibrium and non-equilibrium states
in dynamical systems on the basis of the time average. We show
 numerically  that
  there exists a non-equilibrium non-stationary state in the coupled modified Bernoulli map lattice.
\end{abstract}

\maketitle


\section{Introduction}

In statistical mechanics, it is assumed that macroscopic observables are the result of the time average of microscopic observables. 
 To establish the equilibrium state compatible with thermodynamics in dynamical systems, Boltzmann proposed the ergodicity; in ergodic systems, the time average of an observation function is equal to its space average. Birkhoff proved this assumption for the 
ergodic  probability preserving transformation \cite{Birkhoff1931}. In mathematics, the dynamical system $(X, \B, m, T)$ is ergodic if $m(A)=0$ or $m(A^c)=0$ for all $A \in \B$ satisfying $T^{-1}A=A$ .\par
In thermodynamics, 
if a macroscopic system is in equilibrium, its subsystems 
in the space of position, which are
 macroscopic system,  remain in equilibrium.
However,  the ergodicity proposed by Boltzmann does not provide a equilibrium state in
macroscopic subsystems. Moreover,  macroscopic observables in a non-equilibrium state are intrinsically random, but the randomness of the time average of a microscopic observation function has not been 
studied yet.\par
In this paper, we propose a definition of the equilibrium and non-equilibrium states in dynamical systems on the basis of the time average. Our goal is to determine the measures characterizing a non-equilibrium non-stationary state. 
From the recent progress of the infinite ergodic theory, it is known that the time average of a observation function converges in distribution\cite{Aaronson1997,Thaler1998,Thaler2002,TZ2006}. Moreover, 
distributions of the time average depend on the invariant measure as well as
 the observation function \cite{Akimoto2008}. Therefore, the randomness of 
macroscopic observables in non-equilibrium state can be characterized by that of 
infinite measure systems.\par
This paper is organized as follows. First, distributions
 for the time average of some observation functions are presented according to 
the invariant measure and the observation function. Next, we define
 the equilibrium and non-equilibrium steady states and the non-equilibrium 
non-stationary state.
 Finally, we demonstrate the non-equilibrium non-stationary state using the coupled modified Bernoulli map lattice.

\section{Equilibrium and non-equilibrium states in dynamical systems}

\paragraph{Distributions of time average}
We present distributional limit theorems using the 
modified Bernoulli map $T$ defined by
\begin{equation}
Tx = \left\{
\begin{array}{ll}
x+2^{B-1}x^B \quad &x\in [0,1/2)\\
\\
x-2^{B-1}(1-x)^B &x\in [1/2,1].
\end{array}
\right.
\label{eq:2.1}
\end{equation}
According to \cite{Akimoto2008}, we consider the following time average of the 
observation function $f(x) : [0,1] \rightarrow {\mathbb R}$:
\begin{equation}
\Pr \left(
\frac{1}{a_n} \sum_{k=0}^{n-1} f\circ T^k \leq t \right)
=G(t),
\label{eq:2.2}
\end{equation}
where $a_n$ is regularly varying at $\infty$ with index $\kappa$
 which depends on the invariant measure 
and the observation function $f(x)$.  Universal 
distributions for the time average of some observation functions are summarized
in Table \ref{tab:1}, where the $L^1_{loc,m}$ function with infinite mean 
is written as
\begin{equation}
x^{\alpha}g(x) =O(1), \quad x\rightarrow 0,
\label{eq:2.3}
\end{equation}
\begin{equation}
(1-x)^{\alpha}g(x) =O(1), \quad x\rightarrow 1.
\label{eq:2.4}
\end{equation}
It is worth noting that the time average of the $L^1_{loc,m}$ function with
 infinite mean is 
intrinsically random in the infinite measure case as well as the
 finite measure case.

%


\begin{table}

\label{tab:1}       
\begin{tabular}{llll}
\hline\noalign{\smallskip}
Invariant measure & Observation function $f(x)$ & Distribution $G(t)$ & Exponent $\kappa$ \\
\noalign{\smallskip}\hline\noalign{\smallskip}
Finite ($B<2$)& $L^1(m)$ & {\it Delta} & 1 \\
Finite ($B<2$)& $L_{loc}^1(m)$ with infinite mean &  {\it Stable} & $(2-B)/\alpha$ \\
Infinite ($B\geq 2$)& $L^1(m)$ & {\it Mittag-Leffler} & $1/(B-1)$\\
Infinite ($B\geq 2$)& $L_{loc,m}^1$ with finite mean & {\it Generalized arcsine} & 1\\
Infinite ($B\geq 2$)& $L_{loc,m}^1$ with infinite mean & {\it Stable} & $\alpha/(B-1)+1$ \\
\noalign{\smallskip}\hline
\end{tabular}
\caption{Universal distributions of the time average of the observation function $f(x)$.}
\end{table}

\paragraph{Definition of the equilibrium and non-equilibrium states}
Consider a classical system containing $n$ particles with positions and 
momenta. 
Suppose that $T$ represents the change in positions and momenta of $n$ 
particles during some period of time,\footnote{We do not assume that it is described by a Hamiltonian.} and $X$ is the phase space.
 We call $(X,\B ,m,T)$
 a dynamical system, where $(X, \B, m)$ is a standard $\sigma$-finite 
measure space. Let $\Omega$ be the space of position.
\par
A dynamical system is in 
equilibrium state for the observation function $f(x) : 
X\rightarrow {\mathbb R}$ 
if the following two conditions hold. For all $n$ and partitions 
$\xi =(A_1,\cdots, A_L)$ with 
$\Omega=\cup_{i=1}^L A_i$
\begin{equation}
F(n) \equiv \lim_{N\rightarrow \infty}
\frac{1}{N} \sum_{k=n-N+1}^n f\circ T^k
= \langle f \rangle \equiv \int_X f dm 
\label{eq:3.1}
\end{equation}
and 
\begin{equation}
F_i(n) \equiv \lim_{N\rightarrow \infty}
\frac{1}{N} \sum_{k=n-N+1}^n f_{A_i}\circ T^k
= \langle f \rangle \quad {\rm for~all}~i=1, \cdots , L,  
\label{eq:3.2}
\end{equation}
where $A_i$ is a subsystem divided on the space $\Omega$,
 and $f_{A_i}(x) : X \rightarrow {\mathbb R}$ 
 is the function $f(x)$ restricted to the space 
$A_i$. If equation (\ref{eq:3.2}) holds, we call  
$f(x)$ an intensive function.
\par
A dynamical system is in a non-equilibrium steady state for the observation 
function $f(x)$ 
if the following two conditions hold. For $n$ and partitions 
$\xi$ with 
$\Omega=\cup_{i=1}^L A_i$, equation (\ref{eq:3.1}) is satisfied, 
and $F_i(n)$ exists  and 
$F(n) \ne F_i(n)$ $\forall i=1,\cdots, L$.
\footnote{When the partition is suitable, there exist $i$ and $j$ $(i\ne j)$ such that $F_i(n)=F_j(n)$ for all $n$.}
\par

A dynamical system is in a non-equilibrium non-stationary state for the 
observation function $f(x)$
if  $F_i(n)$ is random for all $n$ and partitions 
$\xi$ with 
$\Omega=\cup_{i=1}^L A_i$,  and there does not exist 
$i$ and $j$ $(i\ne j)$ such that $F_i(n)=F_j(n)$ for all $n$.

\paragraph{Non-equilibrium non-stationary state in a
coupled modified Bernoulli lattice}

We consider a one-dimensional lattice system, where $X=[-1,1]^K$ and 
the space $\Omega$ is considered as the configuration of a lattice, i.e., 
$\Omega =\{1,\cdots ,K\}$. 
Let $x_i(n)$ be a microscopic 
state in the lattice $i$ at time $n$. We suppose that the
time evolution of a microscopic state, $x=(x_1, \cdots, x_K)$, $T : X\rightarrow X$ is given by the coupled map lattice: 
\begin{equation}
\begin{array}{ll}
x_i(n+1) = (1-\epsilon) T_1(x_i(n))+\frac{\epsilon}{2} \{x_{i-1}(n)+
x_{i+1}(n)\} \quad (i=2, \cdots, K-1),\\
\\
x_1(n+1)= (1-\epsilon) T_1(x_1(n)) +\epsilon x_2(n),\\
\\
x_K(n+1)=(1-\epsilon) T_1(x_K(n)) +\epsilon x_{K-1}(n),
\end{array}
\label{eq:3.5}
\end{equation}
where $\epsilon$ is a coupling constant and 
the transformation $T_1$ is defined as\footnote{The transformation $T_1$ around the indifferent fixed point $x=0$ 
is similar to the modified Bernoulli map.}
\begin{equation}
T_{1}x=\left\{
\begin{array}{ll}
-4(1-\epsilon)x+3(1-\epsilon) \quad &x\in [1/2,1],\\
\\
x+(\frac{1}{2}-\epsilon)2^Bx^B &x\in [0,1/2),\\
\\
x-(\frac{1}{2}-\epsilon)2^B(-x)^B &x\in [-1/2,0),\\
\\
-4(1-\epsilon)x-3(1-\epsilon) \quad &x\in [-1,-1/2],
\end{array}
\right.
\label{eq:3.6}
\end{equation}

  In this study, we divide the space 
$\Omega$ into $L$ or $L+1$ subsystems, to be more precise, 
$A_l=\{ (l-1) \lfloor K/L \rfloor +1, \cdots, l \lfloor K/L\rfloor \}$ $(l=1,\cdots ,L)$ and $A_{L+1}=\{L \lfloor K/L\rfloor+1,\cdots ,K\}$ when $K$ is not divisible by $L$, where $\lfloor \cdot \rfloor =\max \{n\in {\mathbb Z}| n\leq \cdot \}$. Then the
macroscopic observables 
$F_l(n)$ $(l=1, \cdots, L)$ are defined by 
\begin{equation}
F_l(n)=\frac{1}{N} \sum_{k=n-N+1}^{n} f_{A_l} (x(k)) =\frac{1}{N} \sum_{k=n-N+1}^{n} \sum_{i=(l-1)K/L+1}^{lK/L+1}f_i(x(k))/(K/L),
\label{eq:3.7}
\end{equation}
where 
\begin{equation}
f_i(x)= \left\{
\begin{array}{ll}
1 \quad &(x_i \geq 0)\\
\\
0 &(x_i <0)
\end{array}
\right.
\quad {\rm for}\quad i=1, \cdots,K.
\label{eq:3.8}
\end{equation}
As shown in Fig. 1,  a homogeneous pattern is clearly observed 
in the case of a finite measure; i.e,, this dynamical system is in an 
equilibrium state for $f(x)=\sum_i f_i (x)/K$. On the other hand, the time evolution of time average $F_l(n)$ is not 
homogeneous but has a complex pattern in the case of an infinite measure; i.e., this dynamical system is in a non-equilibrium non-stationary state for 
$f(x)$.

\begin{figure}[h]
  \includegraphics[height=.5\linewidth, angle=-90]{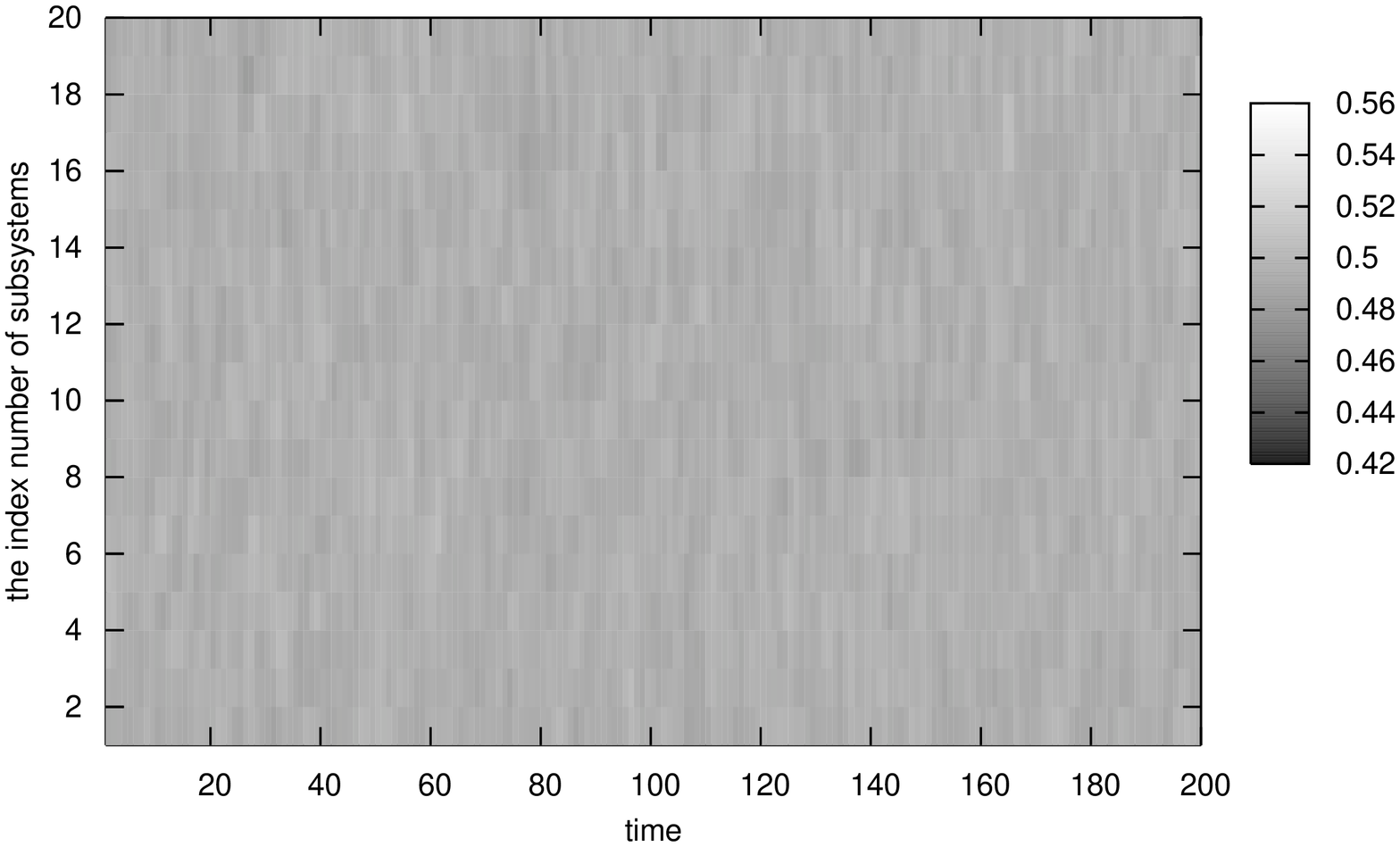}  
\includegraphics[height=.5\linewidth, angle=-90]{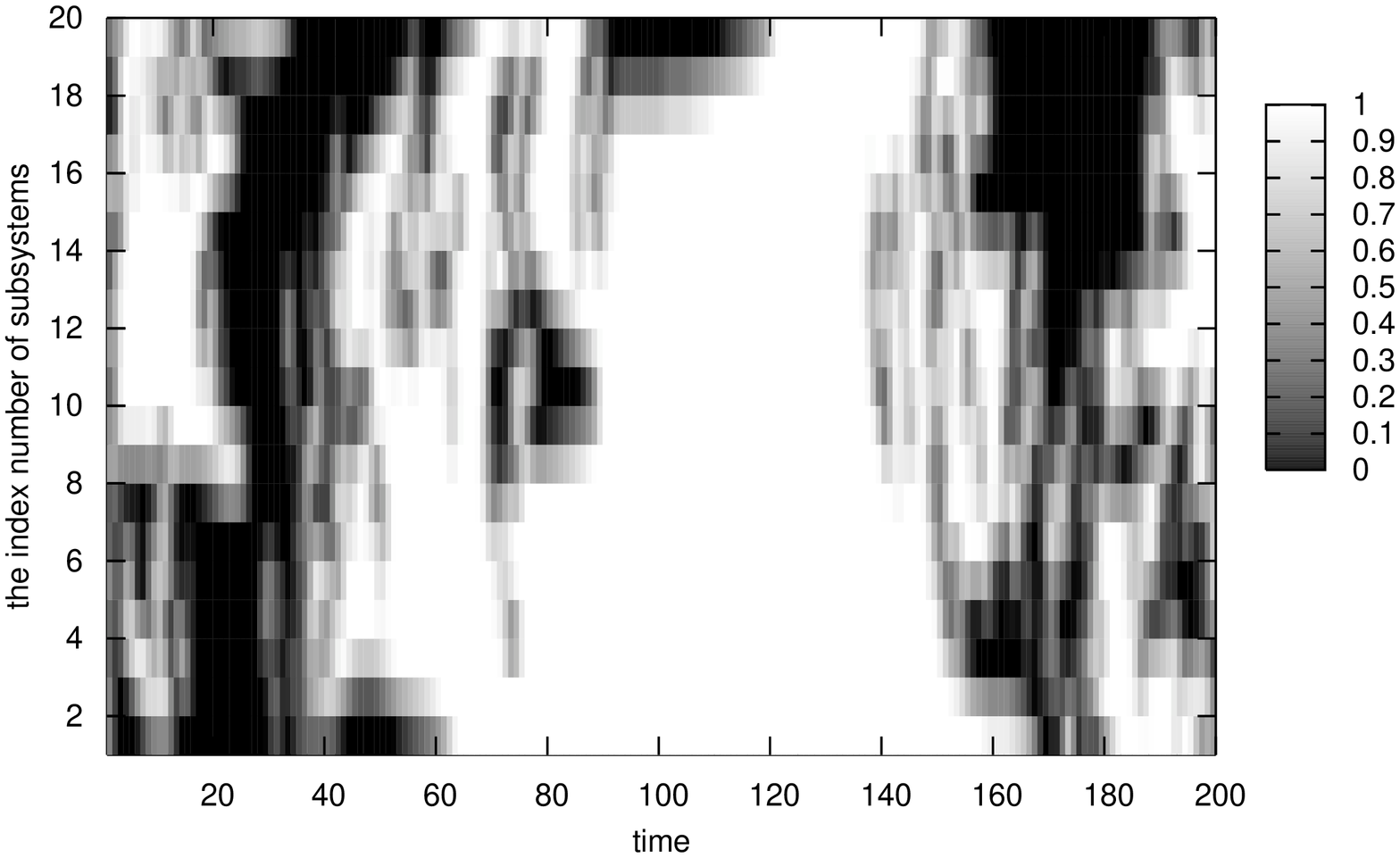}  
\caption{Time evolution of macroscopic observables $F_l(n)$. ($\epsilon=10^{-3}, N=10^4, K=100, L=20$. Left : $B=1.5$, right : $B=3.0$.) The degree of $F_l(n)$ is represented by its thickness.}
\end{figure}

\section{Discussion}
We defined 
the equilibrium and non-equilibrium states in dynamical systems.
In this context,
 the ergodic measure in infinite measure systems 
is one of the measures characterizing the non-equilibrium non-stationary state,
 because macroscopic observables, which
are the result of the time average of  microscopic observables, are 
intrinsically random. In fact, the numerical simulation of the coupled modified Bernoulli map lattice has provided a non-equilibrium non-stationary state
clearly in the infinite measure case.





\bibliographystyle{aipproc}   

\bibliography{akimoto}

\IfFileExists{\jobname.bbl}{}
 {\typeout{}
  \typeout{******************************************}
  \typeout{** Please run "bibtex \jobname" to optain}
  \typeout{** the bibliography and then re-run LaTeX}
  \typeout{** twice to fix the references!}
  \typeout{******************************************}
  \typeout{}
 }

\end{document}